\documentclass{arxiv}
\usepackage[T1]{fontenc}
\usepackage[latin9]{inputenc}
\usepackage{amstext}
\usepackage{graphicx}
\usepackage{multirow}

\title{Optimization of radiation hardness and charge collection of edgeless silicon pixel sensors for photon science}

\author{Jiaguo Zhang$^a$\thanks{Corresponding
author.}~, Damaris Tartarotti Maimone$^b$, David Pennicard$^a$, Milija Sarajlic$^a$, and Heinz Graafsma$^{a}$\\
\llap{$^a$}Deutsches Elektronen-Synchrotron (DESY)\\
  Notkestrasse 85, D-22607 Hamburg, Germany\\
\llap{$^b$}Institute of Physics "Gleb Wataghin", State University of Campinas\\
  Cidade Universitaria Zeferino Vaz - Barao Geraldo, Campinas - SP, 13083-970, Brazil\\
  E-mail: \email{jiaguo.zhang@desy.de}}

\abstract{Recent progress in active-edge technology of silicon sensors enables the development of large-area tiled silicon pixel detectors with small dead space between modules by utilizing edgeless sensors. Such technology has been proven in successful productions of ATLAS and Medipix-based silicon pixel sensors by a few foundries. However, the drawbacks of edgeless sensors are poor radiation hardness for ionizing radiation and non-uniform charge collection by edge pixels. In this work, the radiation hardness of edgeless sensors with different polarities has been investigated using Synopsys TCAD with X-ray radiation-damage parameters implemented. Results show that if no conventional guard ring is present, none of the current designs are able to achieve a high breakdown voltage (typically < 30 V) after irradiation to a dose of $\sim$10 MGy. In addition, a charge-collection model has been developed and was used to calculate the charges collected by the edge pixels of edgeless sensors when illuminated with X-rays. The model takes into account the electric field distribution inside the pixel sensor, the absorption of X-rays, drift and diffusion of electrons and holes, charge sharing effect, and threshold settings in ASICs. It is found that the non-uniform charge collection of edge pixels is caused by the strong bending of electric field and the non-uniformity depends on bias voltage, sensor thickness and distance from active edge to the last pixel ("edge space"). In particular, the last few pixels close to the active edge of the sensor are not sensitive to low-energy X-rays (< 10 keV) especially for sensors with thicker Si and smaller edge space. The results from the model calculation have been compared to measurements and good agreement was obtained. The model can be used to optimize the edge design. From the edge optimization, it is found that in order to guarantee the sensitivity of the last few pixels to low-energy X-rays, the edge space should be kept at least 50\% of the sensor thickness.
}

\keywords{Edgeless sensor; active edge; radiation damage; charge collection; Medipix}

\begin{document}


\section{Introduction}

Large-area hybrid silicon pixel detectors are widely used for imaging experiments at synchrotron radiation sources and Free-Electron Lasers (FELs). Examples are Large Area Medipix-Based Detector Array (LAMBDA) \cite{LAMBDA} at PETRA III, Cornell-SLAC Pixel Array Detector (CSPAD) \cite{CSPAD} at LCLS and Adaptive Gain Integrating Pixel Detector (AGIPD) \cite{AGIPD} at the European XFEL. The large detection area of these detectors is obtained by tiling single hybrid modules together. However, the drawback is the formation of a dead region between individual modules, typically of a few mm. This results in X-ray information missing in this region and may cause further problems during image reconstruction. 

\begin{figure}[htbp]
\small
\centering
\includegraphics[scale=0.27]{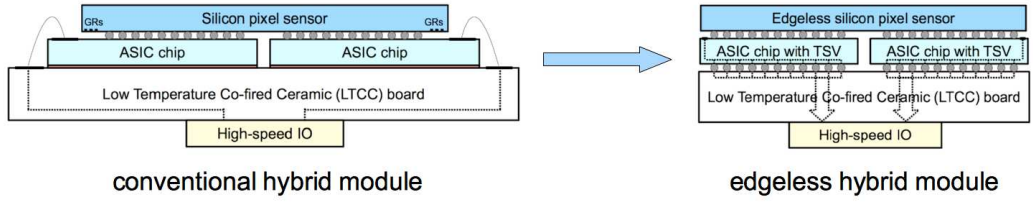}
\caption{From conventional to edgeless hybrid detector module.}
\label{EdgelessDetector}
\end{figure}

The main causes of the dead region from single hybrid detector modules are the guard-ring structure of the sensor, the space occupied by the wire-bonding connection between the ASIC chips and the circuit board. Figure \ref{EdgelessDetector} (left) shows the cross section of a conventional hybrid detector module and its dead-space region. The dead space can be reduced by utilizing edgeless detectors with edgeless sensor replacing conventional silicon sensor, with Through-Silicon-Vias (TSVs) and a Re-Distribution Layer (RDL) implemented into the ASIC chips, and with an integration from ASIC chips to circuit board through a Ball-Grid-Array (BGA). Figure \ref{EdgelessDetector} (right) shows the cross section of an edgeless detector module. At DESY, the edgeless silicon pixel detector based on current LAMBDA detector system is under development.

Recent progress in active-edge techonology of silicon sensors enables the development of edgeless detectors. Such technology has been proven in successful productions of ATLAS and Medipix-based silicon pixel sensors by a few foundries \cite{MBomben, JKCharge}. The key process involves a Deep Reactive Ion Etching (DRIE) to form a trench surrounding the active pixels and a side implantation with the same dopants as backside implant to shield the leakage current from defects located at the edge produced by DRIE \cite{XWu}. The aim of this work is to (1) investigate the radiation hardness of edgeless sensors with active edges using TCAD simulation, (2) understand the charge-collection behavior of edgeless sensors through model calculation, and (3) optimize the edge design for better charge collection by edge pixels and for higher breakdown voltage after high-dose X-ray irradiation.

\section{Investigated structures}

The investigated edgeless sensor is a commercial available product fabricated by VTT (which was later on transferred to its spin-off company Advacam \cite{Advacam}) in a Multi-Project Wafer (MPW) run.

\begin{figure}[htbp]
\small
\centering
\includegraphics[scale=0.2]{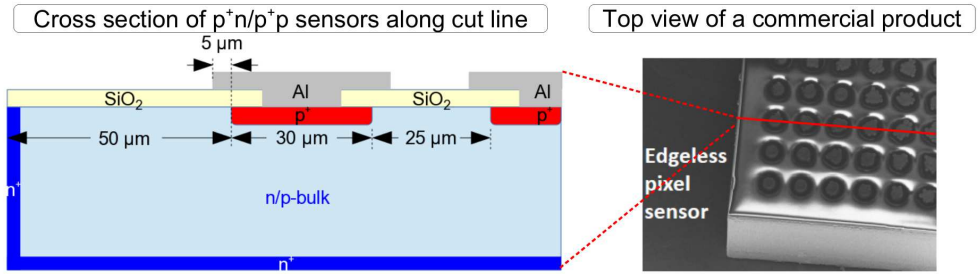}
\caption{Investigated edgeless sensor produced by VTT/Advacam. Left: Cross section of the edgeless sensor along the cut-line through the center of pixels; Right: Top view of the edgeless sensor (picture reproduced from \cite{JKalliopuska}).}
\label{CrossSection}
\end{figure}

Figure \ref{CrossSection} shows the top view of the investigated edgeless sensor and its cross section. The pitch of pixels is 55 $\mu$m, the width/diameter of pixel implant 30 $\mu$m, which leaves a gap of \mbox{25 $\mu$m} between neighbouring pixel implants. The metal overhang, which is the overlap between aluminum plate and implant window of pixel, is 5 $\mu$m. The distance from the implant boundary of the last pixel to the active edge is 50 $\mu$m, also referred to "edge space" or "last pixel-to-edge distance". The thickness of the SiO$_{2}$ layer, extracted from the capacitance of a MOS capacitor produced on the same silicon wafer biased into accumulation, is 700 nm for n-type and 680 nm for p-type silicon. The junction depth of pixel electrode is $\sim$1.2 $\mu$m for both boron and phosphorus dopants, and the depth of p-spray of 0.9 $\mu$m, which were obtained either from process simulations or from a direct measurement with spreading resistance method. The thickness of SiO$_{2}$ and profiles of implants at pixel side are critical parameters for breakdown simulation.

The edgeless sensors with the following polarities have been investigated: p$^{+}$n, p$^{+}$p, n$^{+}$n and n$^{+}$p with either p-spray or p-stop. For sensors with p-stop, the p-stop implant is 5 $\mu$m wide and located in the center of gaps between pixel implants and in the "edge space" region.

\section{Simulation of electrical properties of edgeless sensors}

The active volume and breakdown voltage as function of radiation dose have been simulated with Synopsys TCAD \cite{TCAD}. In the simulation, the following physics models have been implemented: (1) drift and diffusion of carriers, (2) band-gap narrowing, (3) doping dependence and high-field satuaration of carrier mobility, (4) carrier-carrier scattering and mobility degradation at the interface, (5) doping, temperature dependence and electric field enhancement of Shockley-Read-Hall (SRH) recombination, (6) band-to-band tunneling with Hurkx model, (7) avalance process using vanOverstraetenMan model with the gradient of quasi-Fermi potential as driving force, and (8) fixed charges and generation-recombination at the Si-SiO$_{2}$ interface. In addition, Neumann boundary condition was used on top of SiO$_{2}$, which represents a sensor operation in vacuum or dry atmosphere without any influence due to humidity effect. 

The carrier lifetime for both electrons and holes used in the simulation is 1.35 ms, which was extracted from the leakage current of a diode biased into full depletion. The doping concentrations, obtained from the Capacitance-Voltage (CV) measurements on diodes, are $7 \times 10^{11}$ cm$^{-3}$ and $1.1 \times 10^{12}$ cm$^{-3}$ for n-type and p-type silicon, respectively. Before irradiation, the oxide-charge densities are $1.0 \times 10^{10}$ cm$^{-2}$ (n-type), and $3.0 \times 10^{10}$ cm$^{-2}$ (p-type); the surface-recombindation velocities are 1.35 cm/s (n-type) and 3.54 cm/s (p-type). 

The simulations were performed at 20$^{\circ}$C in 2D covering the region from the active edge to pixel-10 (10$^{\textrm{th}}$ pixel counted from the active edge). A large simulation region was chosen so that the simulated boundary cutting through pixel-10, where a Neumann boundary applies, will not affect the electric field distribution close to the sensor edge. Sensors with different thicknesses have been simulated, but results for 300 $\mu$m thick Si will be shown.

\subsection{Active volume}

The sensor active volume is obtained by simulating the potential distribution of edgeless sensors biased at different voltages. 

\begin{figure}[htbp]
\small
\centering
\includegraphics[scale=0.5]{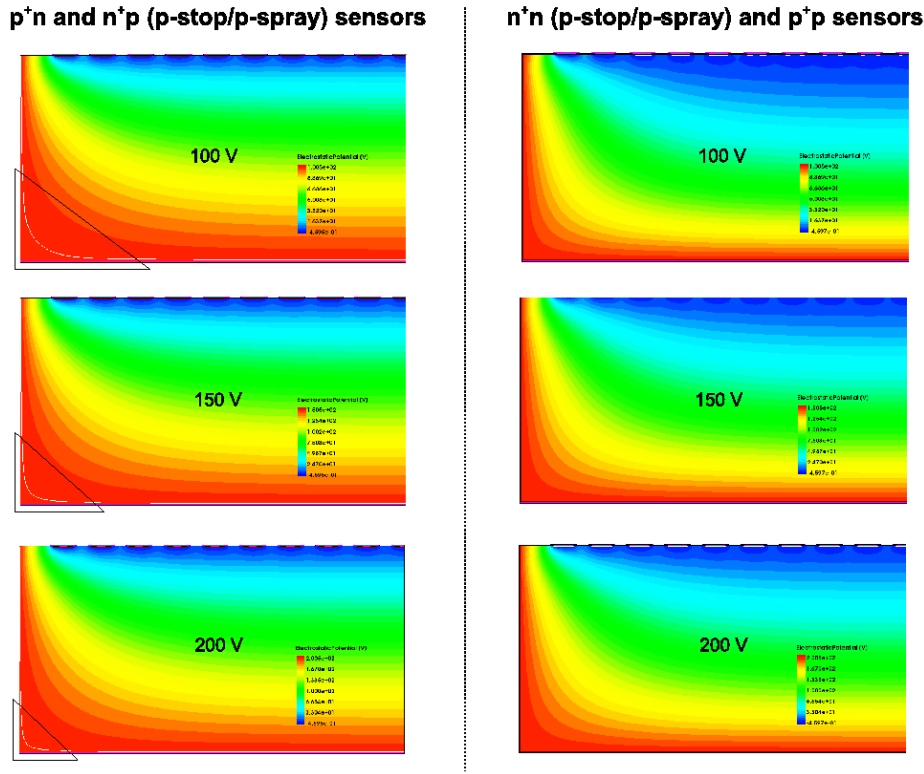}
\caption{The potential distributions at different bias voltages for 300 $\mu$m thick sensors without irradiation. Left: Results for p$^{+}$n and n$^{+}$p (p-stop/p-spray) sensors. Right: Results for n$^{+}$n (p-stop/p-strap) and p$^{+}$p sensors. Note the sign of the bias voltage for different polarities.}
\label{ActiveVolume}
\end{figure}

Figure \ref{ActiveVolume} (left) shows the results for p$^{+}$n and n$^{+}$p (p-stop/p-spray) sensors biased at 100 V, \mbox{150 V} and 200 V (note the sign of the bias voltage for different polarities). The white line in the figure refers to the depletion boundary of the sensor. For these sensors, the depletion starts from the pixel side and extends to the active edge and backside of the sensor, which leaves an "inactive" (non-depleted) region close to the corner between edge implant and backside implant. The volume of the "inactive" region is influenced by bias voltage as seen in the figure, doping concentration and "edge space". The volume shrinks with increasing bias voltage for a given layout and doping concentration. This can change the electric field distribution close to the edge region and result in a bias voltage dependence of charge collection by pixels close to the active edge. To minimize the volume of the "inactive" region and thus eliminate the dependence of charge collection on bias voltage, the bias should be kept much higher than the depletion voltage of the sensor. Typically, a bias voltage of $\sim$150 V above the depletion voltage is expected to be able to remove such influence for a 300 $\mu$m thick silicon sensor.

In figure \ref{ActiveVolume} (right), the potential distributions for n$^{+}$n (p-stop/p-spray) and p$^{+}$p sensors are shown. As the depletion starts from the active edge and the backside electrode, the entire sensor volume will be fully depleted once the depletion reaches the pixel electrodes. This will not leave "inactive" region inside the sensor. Thus, the charge collection by pixels close to the active edge does not strongly depend on bias voltage once the sensor is biased above its depletion voltage.

\subsection{Sensor breakdown and radiation hardness}

One of the main aims of this work is to investigate the breakdown voltage of edgeless sensors as function of X-ray dose. For sensors after X-ray irradiation, oxide charges, interface traps and border traps will be introduced either in the SiO$_{2}$ insulating layer or at the Si-SiO$_{2}$ interface \cite{JZMOS}. In the simulation, oxide charges and charged interface traps were represented as fixed charges at the Si-SiO$_{2}$ interface. In addition, the current generated by the states of interface traps close to the silicon mid-gap was described by surface-recommbination velocity and implemented in TCAD simulation. The details of the input parameters related to X-ray radiation damage can be found in \cite{JZMOS, JS}. The effect of border traps were not simulated due to the lack of quantative characterization.

\begin{figure}[htbp]
\small
\centering
\includegraphics[scale=0.55]{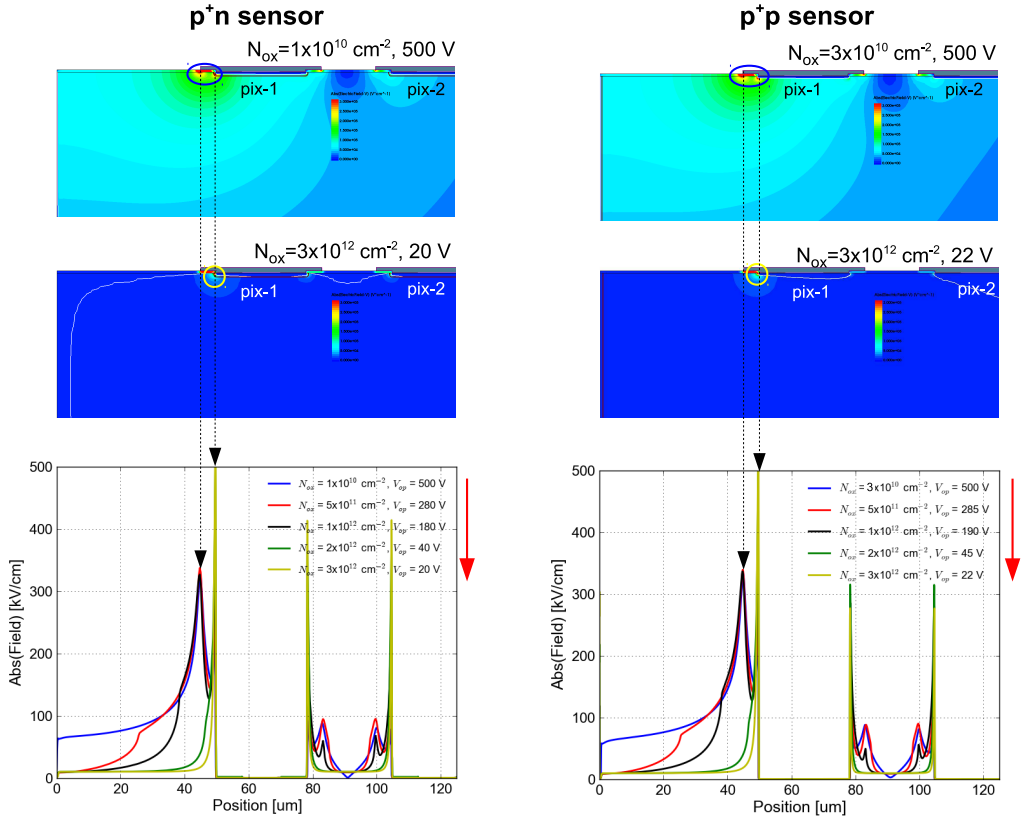}
\caption{Electric field close to the sensor corner at pixel side and its distribution along the cut-line at \mbox{10 nm} below the Si-SiO$_{2}$ interface as function of X-ray irradiation dose represented by fixed charges (\mbox{$N_{ox}=3 \times 10^{12}$ cm$^{-2}$} corresponds to $\sim$10 MGy). Left: p$^{+}$n sensor; Right: p$^{+}$p sensor.}
\label{ElectricFieldPN}
\end{figure}

Figure \ref{ElectricFieldPN} shows the electric field distributions close to the sensor corner at the pixel side (\mbox{50 $\mu$m} depth) for p$^{+}$n and p$^{+}$p sensors. For both sensors, the highest electric field appears at the Si-SiO$_{2}$ interface below the aluminium of the first pixel before irradiation; after high-dose irradiation, the positive charges in the SiO$_{2}$ and charged interface traps induce a layer of electrons accumulating below the Si-SiO$_{2}$ interface, which conducts the high voltage on the backside of the sensor through the active edge to the first pixel and thus results in a high electric field at the junction of the first pixel. In this case, the first pixel behaves like a Current-Collection Ring (CCR) of a conventional sensor and thus breakdown first. It should be noted that, for both sensors, the breakdown votlage is $\sim$500 V before irradiation, and it sharply decreases with X-ray irradiation dose and finally reaches a value of $\sim$20 V at a high dose equivalent to $\sim$10 MGy. The results are consistent with those published in \cite{JSDesign} for a sensor with 700 nm thick SiO$_{2}$ and without any guard rings.

\begin{figure}[htbp]
\small
\centering
\includegraphics[scale=0.55]{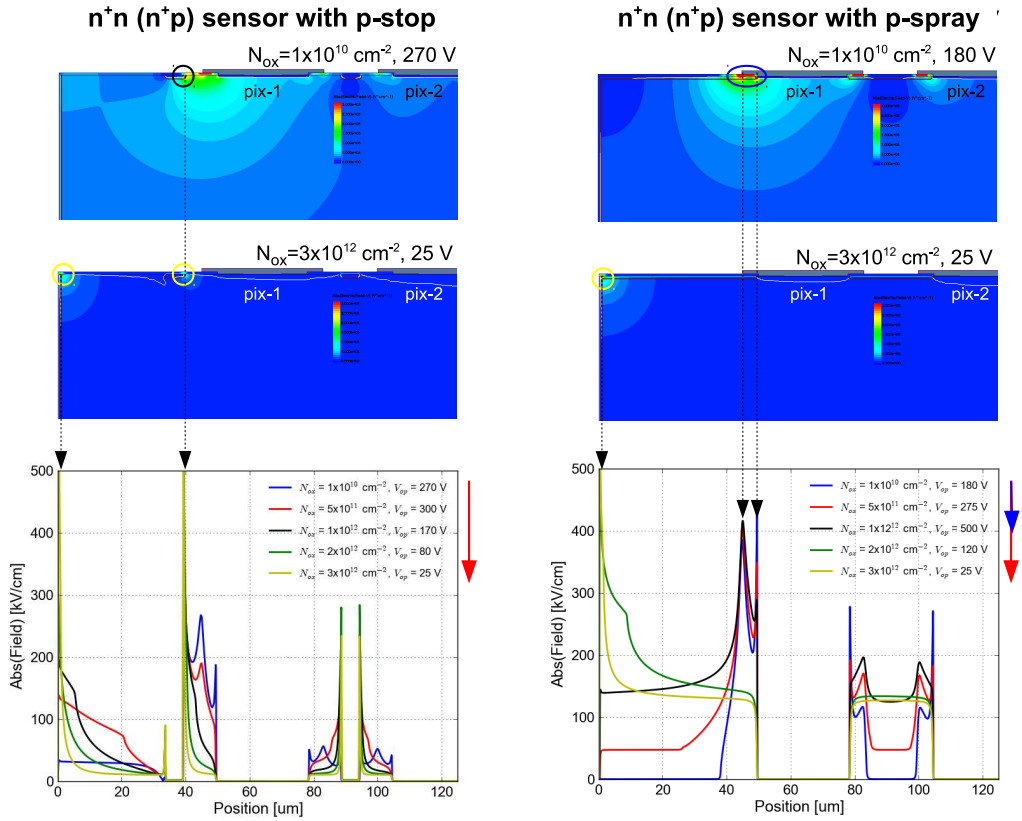}
\caption{Electric field close to the sensor corner at pixel side and its distribution along the cut-line at \mbox{10 nm} below the Si-SiO$_{2}$ interface as function of X-ray irradiation dose represented by fixed charges (\mbox{$N_{ox}=3 \times 10^{12}$ cm$^{-2}$} corresponds to $\sim$10 MGy). Left: n$^{+}$n (and similarly n$^{+}$p) sensor with p-stop; Left: n$^{+}$n (and similarly n$^{+}$p) sensor with p-spray.}
\label{ElectricFieldPP}
\end{figure}

Electric field distributions for n$^{+}$n (n$^{+}$p) sensor with either p-stop or p-spray are shown in figure \ref{ElectricFieldPP}. For n$^{+}$n (n$^{+}$p) sensor with p-stop, the highest electric field is located at the junction of the p-stop between the active edge and the first pixel before irradiation; after irradiation, the high density of electrons from the accumulation layer creates an additional high electric field region at the interface between the edge implant and the accumulation layer and causes a breakdown at the corner of the sensor. For n$^{+}$n (n$^{+}$p) sensor with p-spray, the highest electric fields are located below the aluminum of the first pixel and at the interface between the p-spray layer and the implant of the first pixel before irradiation; after irradiation, a corner breakdown is also observed similar to n$^{+}$n (n$^{+}$p) sensor with p-stop. The breakdown voltage of n$^{+}$n (n$^{+}$p) sensor with p-spray increases with irradiation till the density of fixed charges, $N_{ox}$, larger than the dose of p-spray layer (\mbox{$\sim$1.5$\times$10$^{12}$ cm$^{-2}$}) and then the breakdown voltage starts to decrease. For both n$^{+}$n (n$^{+}$p) sensors with p-stop and p-spray, their breakdown voltage after irradiation is as low as $\sim$25 V at $\sim$10 MGy.


It should be noted that the simulation with drift-diffusion model and fixed charges implemented in gives pessimistic estimation for the breakdown. 

\subsection{Discussion: Towards radiation-hard design}

\begin{figure}[htbp]
\small
\centering
\includegraphics[scale=0.5]{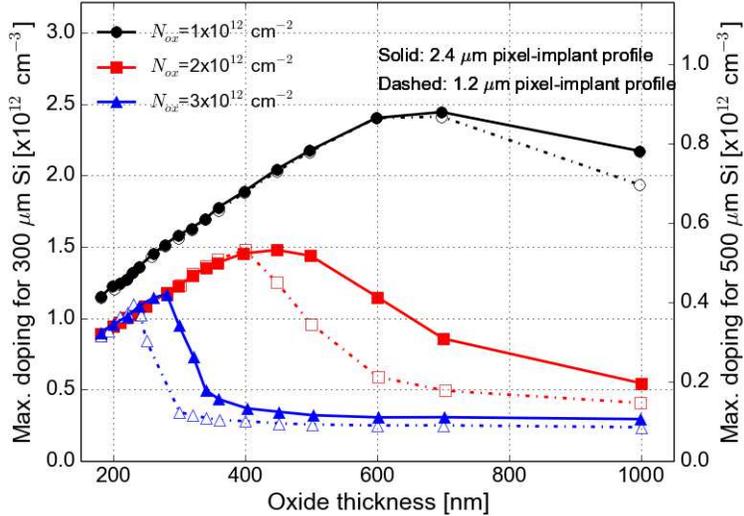}
\caption{Maximal doping in edgeless sensors for radiation hardness purpose. Results re-calculated based on the results from \cite{JSDesign}.}
\label{Breakdown}
\end{figure}

Potentially, there are a few ways to enhance the radiation hardness of edgeless sensors with active edges. The first is to add a guard-ring structure (CCR + floating guard rings) to the region between the active edge and the first pixel. However, this will leave an "inactive" region, because charges generated in some parts of the sensor will be collected by the CCR. The second is to optimize the technological parameters to achieve a high breakdown voltage for a sensor without any guard-ring structure. Here we propose the following methods according to different choices for sensor polarity: 

(1) For p$^{+}$n and p$^{+}$p sensors, the radiation hardness can be achieved by optimizing the oxide thickness below the field plate, junction depth and the doping concentration. In previous work done for the AGIPD sensor, in the case where no floating guard rings are present, the breakdown voltage has been simulated as function of oxide thickness for different junction depth \cite{JSDesign}. Considering the minimal requirement for edgeless sensor operation is that the operating voltage, $V_{op}$, should be larger than the depletion voltage, $V_{dep}$, but smaller than the breakdown voltage, $V_{bd}$, the maximal doping concentration $N_{d}$ can be calculated for silicon sensors with different thicknesses according to  $V_{dep} = \frac{q_{0}T_{si}^{2}N_{d}}{2\varepsilon_{0}\varepsilon_{si}} - V_{bi} \leq V_{op} \leq V_{bd}$. Figure \ref{Breakdown} shows the calculated results on the maximal doping concentration required by 300 $\mu$m and 500 $\mu$m thick sensors as function of technological parameters in order to achieve a breakdown voltage larger than the depletion voltage. It can be seen that, for example, by keeping the doping concentration less than $1.2 \times 10^{12}$ cm$^{-3}$ for 300 $\mu$m Si and $4.0 \times 10^{11}$ cm$^{-3}$ for 500 $\mu$m Si, the breakdown voltage can be higher than the depletion voltage for a sensor with 250 nm thick SiO$_{2}$ and junction depth of 2.4 $\mu$m. Therefore, careful selection of technological parameters makes the radiation hardness of p$^{+}$n and p$^{+}$p edgeless sensors possible.

(2) For the n$^{+}$n and n$^{+}$p sensors with p-spray layer, it has been seen that the breakdown voltage increases as function of irradiation at low doses. This indicates that an optimized/increased dose of p-spray layer can improve the radiation hardness of edgeless sensor to a higher irradiation dose. However, the drawback by increasing the p-spray dose is the reduction of breakdown voltage before irradiation as the increased boron concentration increases the electric field at the interface between the p-spray layer and the pixel implant at a fixed bias voltage. To balance this, the p-spray dose can be chosen as half of the saturation value of the density of fixed charges. Thus, the density of fixed charges has to be characterized as function of dose from the test structures produced by the same technology from the vendor. Another option is to use a p-spray dose same as the saturation value of the density of fixed charges and perform a pre-irradiation to the sensors to certain dose level in order to reduce the effect from the high-dose p-spray. The second needs more investigations before the sensor works.

\section{Modeling charge-collection behavior of edgeless sensors}

The charge collection of edgeless sensors has been widely investigated \cite{JKCharge, RBates}. It has been commonly observed that the charge collection by edge pixels are highly non-uniform. To understand the charge-collection behavior of edgeless sensors, we have developed a model. The model is used to understand the measurement results from thin sensors and predict the performance for thick silicon sensors.

\subsection{Model development}

The model is based a Finite Volume Method (FVM), which segments the sensor into finite volumes. The model considers the following main processes: (1) Photon absorption in silicon sensor and generation of electron-hole pairs, (2) drift of free carriers along electric field and lateral diffusion, and (3) carrier collection by electrodes and signal processing in readout electronics.

In the first process, X-ray photons are absorbed in silicon sensor through photoelectric effect, which is the dominant effect for photon energy less than 20 keV. The probability of photon absorption in each finite volume is given by $1/ \lambda (E_{xray}) \cdot \textrm{exp} [- y/ \lambda (E_{xray})]$, where $\lambda (E_{xray})$ is the attenuation length of X-ray photons with energy $E_{xray}$ and $y$ the distance from the point where the photon is absorbed to the silicon surface where X-rays enter. Once the photon is absorbed, a charge cloud consisting of electrons and holes are created. The number of electron-hole pairs inside the charge cloud is given by the X-ray energy divided by 3.6 eV, which is the mean energy needed to generate one pair of electron and hole in silicon.

In the second process, the electrons and holes drift to either pixel or backside/edge electrode, depending on the sensor polarity. The drifting path of the center of mass of electrons and holes follows the eletric field lines (vector) obtained from TCAD simulation. Then, the drifting time of carriers to pixels is calculated by $t = \int _{x,y} \frac{\textrm{d}y}{\mu \cdot E(x,y)}$, with $\mu$ the carrier mobilty and $E(x,y)$ the electric field. The size of charge cloud when carriers reach pixels is given by the lateral diffusion of carrers: $\sigma = \sqrt{2D\cdot t}$, with $D$ the diffusion coefficient.

Finally, the carriers are collected by different pixels, depending upon their locations when they reach the pixel side. Results from the model calculation can be output either as number of photons or charges depending on whether the ASIC chip is operated in photon-counting or charge-integrating mode. In addtion, a threshold $E_{thr}$ can be set in simulation so that results can be compared to measurements with a certain threshold. The final spectrum can be deconvoluted with the profile of the beam spot for a beam scan experiment.


\subsection{Comparison to measurements}

\begin{figure}[htbp]
\small
\centering
\includegraphics[scale=0.15]{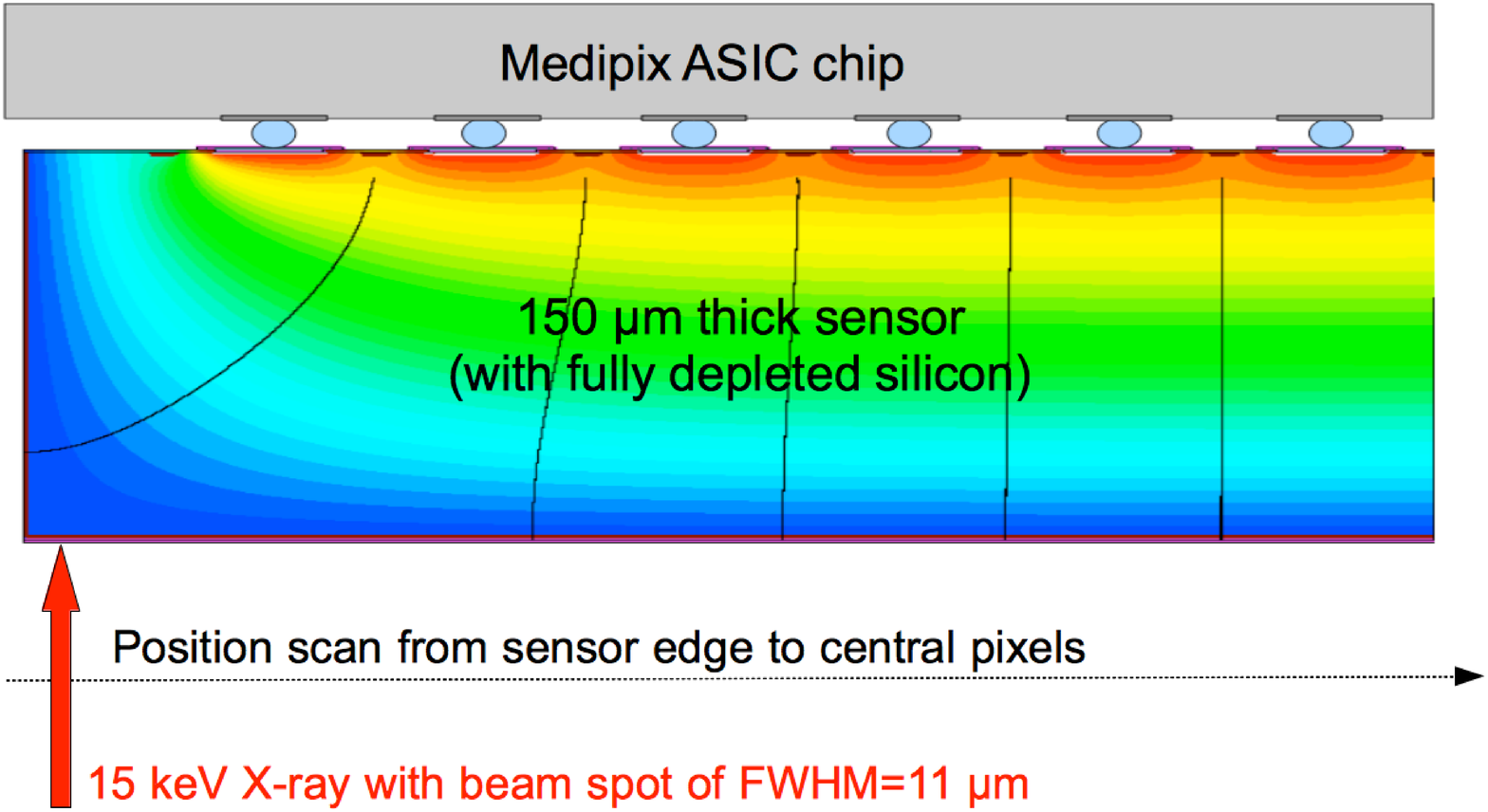}
\includegraphics[scale=0.39]{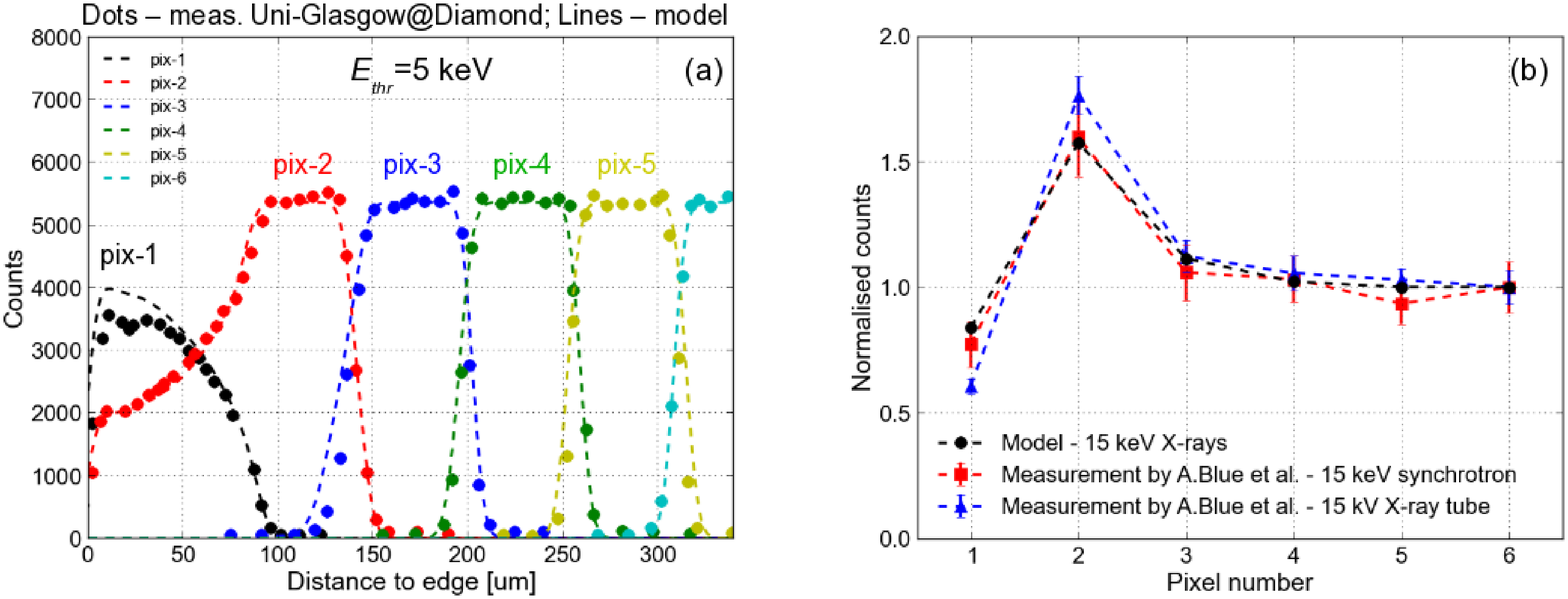}
\caption{Comparison between model calculation and measurements. Top: Measurement scheme and electric field lines inside the sensor; Bottom left: Counts as function of beam position; Bottom right: Normalised counts for edge pixels.}
\label{Comparison}
\end{figure}

Results from model calculation have been compared to measurements in a beam scan experiment done at the Diamond Light Source. The X-ray beam with energy of 15 keV and a spot of FWHM=\mbox{11 $\mu$m} was shot into the backside (non-pixel side) of 150 $\mu$m thick edgeless sensor coupled to a Medipix chip \cite{RBates}. The counts by edge pixels were measured with a threshold setting at 5 keV in Medipix chip as function of distance from the X-ray beam to the edge of the sensor. On top of figure \ref{Comparison}, the measurement scheme is shown.

The comparison between model calculation and measurements is shown on the bottom of figure \ref{Comparison}. A good agreement was obtained. The counts for different edge pixels as function of distance from X-ray beam to the sensor edge is shown in the left figure: Non-uniform charge collection by pixels close to the sensor edge is observed. The non-uniformity of charge collection is explained by the bending of electric field close to the sensor edge: Figure \ref{Comparison} (top) shows the distributions of electric field lines ending at the middle of pixel gaps (10 $\mu$m below the Si-SiO$_{2}$ interface). By integrating the individual distributions for each pixel, the total counts in the scan for each pixels can be obtained, as shown in figure \ref{Comparison} (right). The counts for the edge pixels have been normalised to the count of a central pixel. The result corresponds to the counting behavior of edge pixels in a flat-field image. For 150 $\mu$m thick sensor, the last two pixels count differently compared to the other pixels.

\subsection{Prediction}

In photon science application with hard X-rays, thick silicon sensors are preferred and commonly used in order to obtain a good quantum efficiency. With this developed model, it is possible to predict the charge-collection behavior for thick edgeless silicon sensors.

\begin{figure}[htbp]
\small
\centering
\includegraphics[scale=0.60]{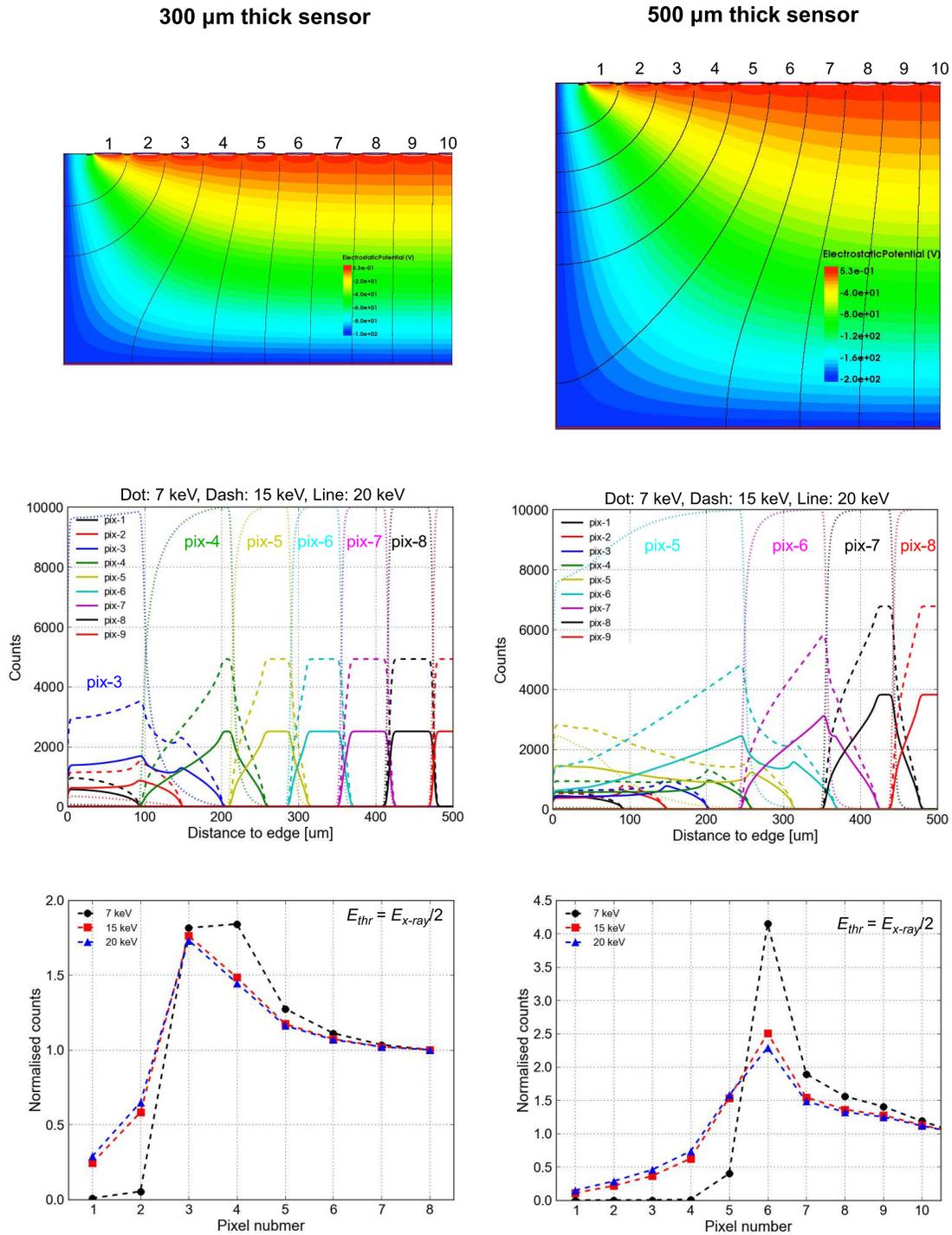}
\caption{Prediction of charge-collection behavior for 300 $\mu$m and 500 $\mu$m thick edgeless sensors with 50 $\mu$m "edge space" from model calculation. Top row: Electric field lines; Middle row: Counts as function of beam position in a beam scan experiment; Bottom row: Normalised counts for edge pixels.}
\label{Prediction}
\end{figure}

Figure \ref{Prediction} shows the electric field lines inside 300 $\mu$m and 500 $\mu$m thick edgeless silicon sensors with 50 $\mu$m "edge space", counts as function of beam position in a beam scan experiment and their normalised counts for edge pixels. It is found that the results depend on the thickness of the sensor and the energy of X-rays. approximately 7 pixels for 300 $\mu$m Si and 10 pixels for 500 $\mu$m Si close to the edge are influenced due to the bending of electric field caused by edge implantation. In addition, for low-energy X-rays ($E_{xray} < 10$ keV), from two to four pixels cannot effectively respond to photons for 300 $\mu$m and 500 $\mu$m thick sensors, respectively.

Therefore, in order to maintain the possibilty of edge pixels to see low-energy photons, we propose the distance from the last pixel to the active edge should be kept at a distance at least 50\% of the sensor thickness.

\section{Summary and outlook}

In this work , radiation hardness of edgeless sensor with active edges has been investigated through TCAD simulation. Results indicate poor radiation hardness for current designs. To improve the radiation hardness, a few methods have been discussed for different sensor polarity choices. In addition, a model has been developed, which makes it possible to reproduce the measurement results at a beam scan experiment with X-rays and predict the charge-collection behavior for different sensor layouts. Finally, the distance from the last pixel to active edge is proposed based on the model calculation in order to obtain a better sensitivity to low-energy photons for edgeless sensors.

\acknowledgments

        The work was performed within the LAMBDA project. J. Zhang would like to thank J. Kalliopuska of Advacam co. for providing process information for TCAD simulation, and D. Maneuski of Glasgow University sharing the measurement data taking at the Diamond Light Source for comparison with the model calculation.

\end{document}